\documentclass[aps,prl,twocolumn,showpacs]{revtex4}
\usepackage{graphics,graphicx}
\usepackage{color}
\usepackage{wrapfig}
\usepackage{enumerate}
\usepackage{amssymb,amsmath}
\usepackage{bm}

\begin{document}
\unitlength = 1mm
\title{Structure of magnetic order in Pauli limited unconventional superconductors}

\author{Yasuyuki~Kato$^1$,  C.~D.~Batista$^1$, I.~Vekhter$^2$}

\affiliation{$^1$Theoretical Division, T-4 and CNLS, Los Alamos National Laboratory, Los Alamos, NM 87545}
\affiliation{$^2$Department of Physics and Astronomy,
             Louisiana State University, Baton Rouge, Louisiana, 70803, USA}

\date{\today}
\pacs{xxx}
\begin{abstract}
We analyse the magnetic structure of the antiferromagnetic order induced in
Pauli limited $d$-wave superconductors by  Zeeman coupling to magnetic field.
We determine the phase diagram in the $H$-$T$ plane, and find that the magnetic phase, which is stabilized
at low temperatures and just below the upper critical field,
can have two realizations depending primarily on the shape of the underlying Fermi surface. The double-${\bm Q}$ magnetic ordering 
may persist over the entire coexistence range. 
Alternatively, there may exist a weak first order transition from a double-${\bm Q}$ structure at lower fields to a single-${\bm Q}$ modulation at higher fields. 
Together with the calculations of the NMR line-shape these results suggest the second scenario as a serious candidate for describing the superconducting state of CeCoIn$_5$.
\end{abstract}
\maketitle

{\it Introduction.} Understanding of the emergence and stability of competing orders in correlated systems has been a major focus of research.  Static long-range order changes the excitation spectrum of  itinerant systems, and the resulting  energy gain determines which instability dominates.  A single order parameter appears in simple cases, but  electron-electron interactions can also favor several distinct phases in many materials. The competition between different 
ordering phenomena make these phases sensitive to applied pressure, chemical doping and magnetic field.

Heavy fermion CeCoIn$_5$ presents one of the most prominent and puzzling examples of such complex behavior. At ambient pressure it is a very clean singlet $d$-wave superconductor with lines of nodes in the gap function. Upon doping, superconductivity coexists with and then is pre-empted by an antiferromagnetic (AFM) order. The isostructural CeRhIn$_5$ is an AFM metal. At low temperatures and fields, $H$, just below the Pauli limited upper critical field $H_{c2}$, CeCoIn$_5$ enters a thermodynamic phase that was initially conjectured~\cite{ABianchi:2003a} to be the first realization of the  Fulde-Ferrell-Larkin-Ovchinnikov (FFLO) state, where the
superconducting (SC) order parameter oscillates in real space  with a length scale proportional to the ratio between $H$ and the Fermi velocity.  However, experiments showed that in this superconductivity  coexists with static long-range incommensurate AFM order~\cite{VMitrovic:2006,BLYoung:2007,MKenzelmann:2008,Koutroulakis10,MKenzelmann:2010,EBlackburn:2010}, even though no AFM order is found upon suppression of SC by doping, magnetic field, or pressure~\cite{ABianchi:2003b,EDBauer:2005,FRonning:2006,CFMiclea:2006}. Incommensurability and small ordered moment are consistent with itinerant, or spin-density wave  (SDW) magnetism, which normally competes for the electronic states with superconductivity.  The inescapable conclusion that in CeCoIn$_5$ superconductivity enables antiferromagnetism challenged our views. 

The existing theories fall broadly into two categories. The first assumes that the spatial variation of the SC order under applied field (due to the FFLO modulation) yields a higher single-particle density of states (DOS)  in the regions where superconductivity is suppressed,  and nucleates a SDW~\cite{DFAgterberg:2009,YYanase:2009,Yanase2011,Yanase2011a}. However, single NMR line shape indicates homogeneous magnetism~\cite{Koutroulakis10,Kumagai11}, and, together with independence of the SDW wave vector on the field~\cite{MKenzelmann:2010},  suggests that the origin of the SDW instability is not related to this modulation (although the FFLO state may still exist~\cite{Kumagai11}). The second category investigates how a strong Pauli limiting in a nodal superconductor may promote a uniform SDW state~\cite{AAperis:2008,AAperis:2010,RIkeda:2010,KSuzuki:2011,Kato11}. We recently showed~\cite{Kato11} that a two-dimensional $d$-wave superconductor under a Zeeman (paramagnetic) field is {\em generically} unstable towards formation of an SDW at the  wave vectors $\pm {\bm Q}_{\pm}$ connecting the opposite nodes of the SC order parameter. The instability is due to nearly perfect nesting of field-induced pockets of Bogoliubov quasiparticles, and explains the incommensurate and field-independent SDW wave vector, as well as the direction of the local moment normal both to the applied field and to the structural layers.

The main challenge to these theories is the structure of the SDW state. Very generally, if the instability originates from  magnetic scattering of quasiparticles between the nodal regions,  the staggered magnetization, $m_{\bm Q}$, of a $d$-wave superconductor has Fourier components for  the four wave vectors  $\pm {\bm Q}_{\pm}$  connecting
the ``nested'' pairs of nodal points, i.e. along each of the two orthogonal directions, $110$ and $1\bar{1}0$. We term this structure double-${\bm Q}$ (2Q). The 2Q SDW gaps all four  pockets of  Bogoliubov quasiparticles thus leading to the greatest energy gain. Therefore, it is always more stable for weak-coupling and hence lower fields~\cite{Kato11}. However, the NMR lineshape~\cite{Koutroulakis10,Kumagai11} is consistent solely with a single-${\bm Q}$ (1Q) SDW modulation. We conjectured previously that intermediate or strong-coupling may stabilize 1Q state, but, to our knowledge, no resolution of this discrepancy, which is generic to the theories of SDW instability of the nodal superconductors, has been offered.

Below we show that  interference between Bogoliubov quasiparticles from the neighborhood of different pockets is the relevant factor
that determines the magnetic structure.  The phase diagram contains a transition line between 2Q and 1Q magnetic orders for a range of parameters in the strong-coupling limit. This transition has weak thermodynamic signatures (such as the specific heat anomaly) but manifests itself very clearly in the change of the NMR lineshape. Unambiguous observation of such a change would strongly favor the current theory for the emergence of the SDW order in CeCoIn$_5$.

\begin{figure}[t]
\hspace{-0.7cm}
\includegraphics[angle=0,width=8.5cm,trim= 30 35 360 50,clip]{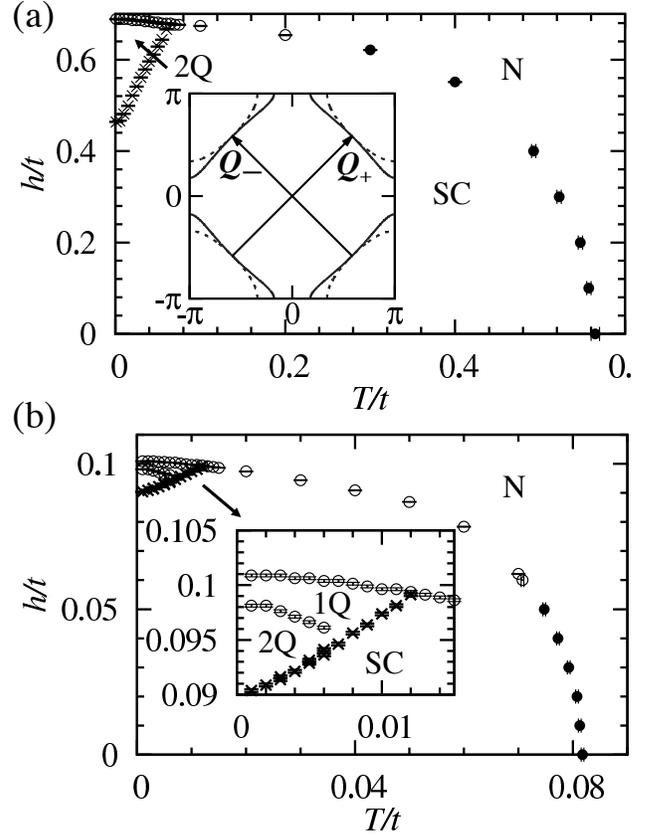}
\caption{
Thermodynamic phase diagrams for
(a) $t'/t=0$, $V/t=3$, $J/t=4$, and
(b) $t'/t=0.2$, $V/t=0.8$, $J/t=2.4$.
Open and closed circles represent first and second order phase transition, respectively.
Crosses represent the transition from single-${\bm Q}$ or double-${\bm Q}$ to uniform superconductivity
for $L=432$.
The inset of (a) shows the nesting vectors ${\bm Q}_{\pm}$ and the Fermi surfaces at $t'/t=0$ (dashed line) and $t'/t=0.2$ (solid line),
with $V/t=0$, and $J/t=0$.}
\label{fig:PDs}
\end{figure}

{\it Model}:
We consider a mean-field Hamiltonian for a $d$-wave superconductor under Zeeman
magnetic field,
\begin{eqnarray}
\mathcal{H} &=& \mathcal{H}_{\rm BCS}+\mathcal{H}_{\rm M}+\frac{N|\Delta_0|^2}{V}+2NJ ( m^2_{{\bm Q}_+} + m^2_{{\bm Q}_-} ),
\nonumber\\
\mathcal{H}_{\rm BCS} &=&\sum_{\bm k, \sigma} \left( \epsilon_{\bm k} - h\sigma \right) c^{\dag}_{{\bm k}\sigma}c_{{\bm k}\sigma}
-\sum_{\bm k}(
	\Delta_{\bm k} c^{\dag}_{{\bm k} \uparrow} c^{\dag}_{-{\bm k} \downarrow}
	+ {\rm {H.c.}})
\nonumber\\
\mathcal{H}_{\rm M}&=&
\!\!\! -J  \!\!\!\!\! \sum_{{\bm k}, \nu=\{\pm\}}  \!\!\!\!\!  ( m_{{\bm Q}_{\nu}}
  c^{\dag}_{\bm {k-Q}_{\nu}\uparrow}c_{{\bm k}\downarrow}
+ m_{{\bar {\bm Q}}_{\nu}}
c^{\dag}_{\bm {k+Q}_{\nu}\uparrow}c_{{\bm k}\downarrow}
	+ {\rm {H.c.}}
)\,.
\nonumber
\end{eqnarray}
Here
$\epsilon_{\bm k} = 2t\left( \cos{k_x} +\cos{k_y}\right)+4t'\cos{k_x}\cos{k_y}-\mu$ is the band energy,
$h=g\mu_B H/2$ is the Zeeman splitting , $N$ is the number of lattice sites, and
 $\Delta_{\bm k} = \Delta_0 \left( \cos{k_x} - \cos{k_y} \right)$ is the superconducting order parameter.
 $J$ denotes magnetic interaction, and
we explicitly separated the four (equal in magnitude) wave-vectors  $\pm \bm Q_\nu$: $\nu=\pm$ indicates a different pair of
"nested" 
gap nodes, 
while ${\bm Q_\nu}$ and ${\overline{\bm Q}_\nu}=-{\bm Q_\nu}$
 are the two vectors connecting a given pair, see inset of Fig.\ref{fig:PDs}(a). Below we select ${\bm Q}_{\pm}=( 7\pi/6,\pm 7\pi/6)$ by fixing the chemical potential to $\mu=\epsilon_{{\bm Q}_+/2}$.
The self-consistency equations are 
\begin{eqnarray}
\Delta_0 &=& \frac{V}{N}\sum_{\bm k}\left( \cos{k_x} - \cos{k_y} \right)\left\langle c^{\dag}_{{\bm k}\uparrow}c^{\dag}_{-{\bm k}\downarrow}\right\rangle ,
\label{eq:sceq1}\\
m_{{\bm Q}}&=&\frac{1}{N}\sum_{\bm k}\left\langle c^{\dag}_{{\bm k} +{\bm Q}\uparrow}c_{\bm k\downarrow}  \right\rangle,
\label{eq:sceq2}
\end{eqnarray}
where
$\langle \cdots \rangle$ denotes the thermodynamic average for $\mathcal{H}$, and $V$ is the pairing interaction in the $d$-wave channel.

The four order parameter components $m_{\bm  Q}$ lead to the following real space distribution of magnetic moments
\begin{eqnarray}
m^{x}_{\bm r} &=&\frac{1}{2}\sum_{\nu=\{\pm\}}
	\left[
		m_{ {\bm Q}_{\nu}}  e^{i {\bm Q}_{\nu}\cdot{\bm r}}   +m_{\overline{\bm Q}_\nu} e^{-i {\bm Q}_{\nu}\cdot{\bm r}}
	\right],
	\nonumber\\
m^{y}_{\bm r} &=&\frac{i}{2}\sum_{\nu=\{\pm\}}
	\left[
		m_{ {\bm Q}_{\nu}}  e^{i {\bm Q}_{\nu}\cdot{\bm r}}   - m_{\overline{\bm Q}_\nu} e^{-i {\bm Q}_{\nu}\cdot{\bm r}}
	\right].
\end{eqnarray}
The easy axis in CeCoIn$_5$ is perpendicular to the layers~\cite{Curro2009}, so the field in the plane clearly favors
the staggered moment along the easy direction, in agreement with experiment~\cite{Kenzelmann08,Curro2009}. We therefore
assume such collinear spin ordering,
$|m_{ {\bm Q}_{\nu}}|=|m_{ \overline{ {\bm Q}}_\nu }|$,
and choose the easy axis as the spin $x$-direction (crystal $c$-axis),
$m_{ {\bm Q}_{\nu}}=m^*_{ \overline{ {\bm Q}}_\nu}$.  There is a zero phase mode for each value of $\nu$ due to  the incommensurate nature of the ordered phase. Without loss of generality
we choose the amplitudes $m_{\bm  Q}$ to be real numbers, i.e., we fix both phases at zero.
Two possible magnetic structures can be stabilized under these conditions.
The 1Q structure, in which only one of the two amplitudes, $m_{ {\bm Q}_{\nu}}$, is finite, corresponds to a SDW that is modulated along the direction parallel to the corresponding ${\bm  Q}_{\nu}$ vector.
The 2Q structure , $|m_{{\bm Q}_{+}} | = |m_{{\bm Q}_{-}} | $,  is modulated along both 
$110$ and $1\bar{1}0$.

\begin{figure}[!htpb]
\includegraphics[angle=0,width=8.5cm,trim= 210 40 540 50,clip]{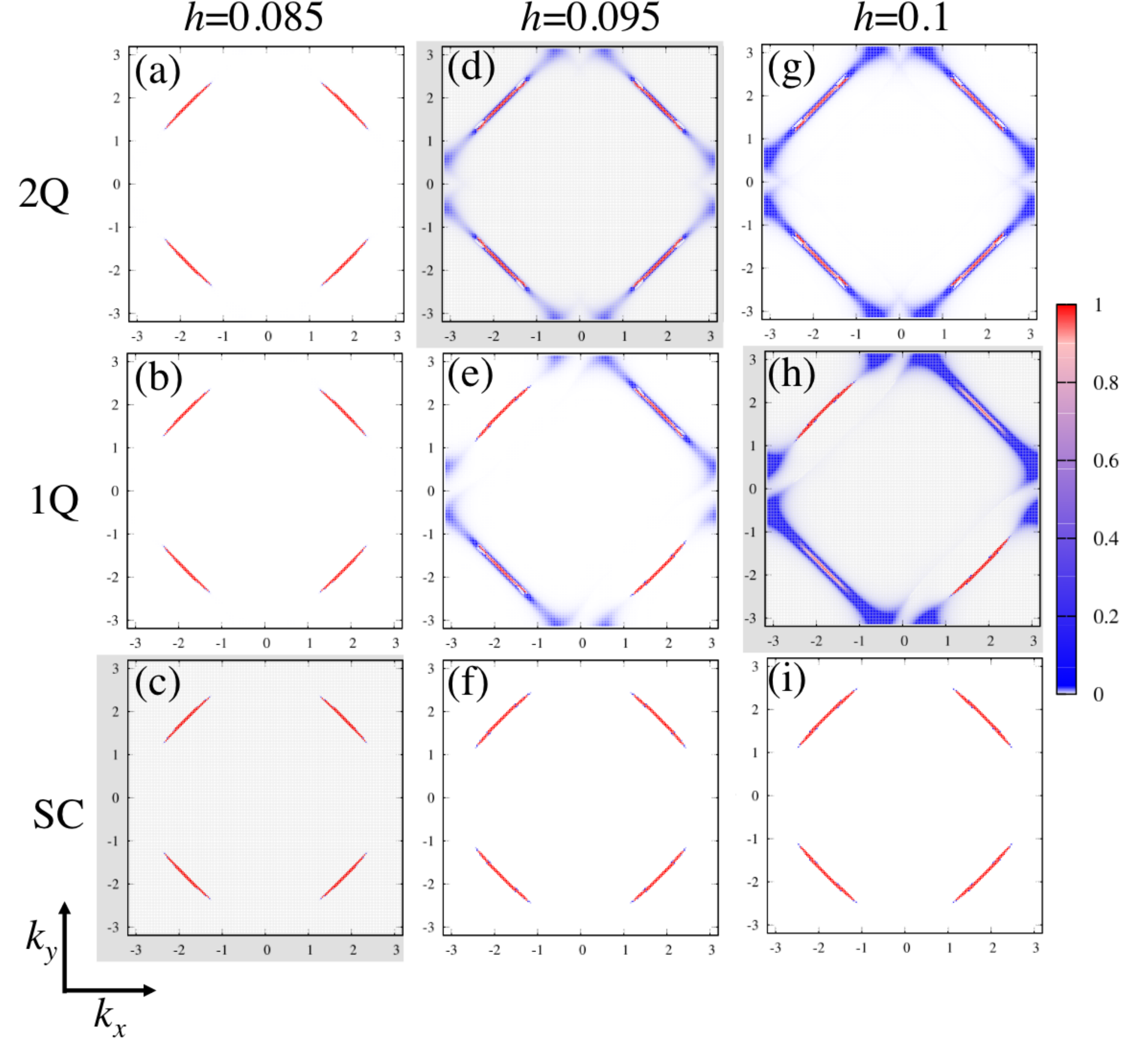}
\caption{
	Density of Bogolyubov quasi-particles $\langle \gamma^{\dagger}_{{\bm k}\uparrow} \gamma^{\;}_{{\bm k}\uparrow} \rangle $ for $t'/t=0.2$, $V/t=0.8$, $J/t=2.4$, and $T/t=10^{-4}$ with $L=864$.
	Shaded panels (c), (d), and (h) correspond to the lowest free energy solution of the self-consistent equations \eqref{eq:sceq2} at a given magnetic field value.
	}
\label{fig:nB}
\end{figure}

By numerically solving 
Eqs. (\ref{eq:sceq1})-(\ref{eq:sceq2}) on a square lattice of linear size $L\sim 864$ and
for different values of band parameters and coupling constants,
we found  two distinct types of phase diagrams that are shown in Figs.~\ref{fig:nB}.
Both of them show the 2Q ordering at low temperatures and moderately high fields.
However, while  this phase persists throughout the whole magnetic region for some parameter values,
Fig.~\ref{fig:PDs}(a), in other cases the SDW phase exhibits a transition between
1Q and  2Q structures, see Fig.~\ref{fig:PDs}(b).
The transition between uniform SC and normal (N) phases is the same in both situations: second order at low fields/high temperatures, and first order at high fields/low temperatures; this behavior is generic for two-dimensional paramagnetically limited superconductors.
The transition from the normal to the magnetically ordered SC phase remains of first order, while the onset of SDW in the SC phase occurs via a second order transition. Finally, a weak first order phase transition line separates the 1Q and 2Q phases.
Since neutron scattering \cite{Kenzelmann08} and NMR measurements \cite{Koutroulakis10}  of  CeCoIn$_5$
clearly indicate the presence of a 1Q phase at high fields, this latter case is relevant~\cite{Kato11}, and we now discuss the origin and the properties of the 1Q/2Q transition in detail.

Recall that in the  weak-coupling (low field) regime, the SDW order arises from nesting between opposite  pockets of Bogoliubov quasiparticles (eigenstates of ${\cal H}_0+{\cal H}_{\rm BCS}$)~\cite{Kato11} generated by the Zeeman term in a nodal superconductor~\cite{Yang98}. The stability of the 2Q  phase arises from the decoupling of the mean field equations \eqref{eq:sceq2} for the four amplitudes $m_{\bm  Q}$, which requires the same amplitude $|m_{\bm  Q}|$ of all the components to fully gap the four nodal pockets. Numerical solutions for all
sets of parameters  always yield the 2Q order at the lowest fields, in agreement with this argument.

The lowest order interaction term  of a  Ginzburg-Landau expansion of the free energy is $|m_{{\bm Q}_+}|^2|m_{{\bm Q}_-}|^2$. Diagrammatically, the coefficient of this term involves a product of four propagators of the  Bogoliubov quasiparticles at momenta separated by $\pm{\bm Q_\pm}$, with the main contribution when all four momenta are near the Fermi surface pockets.
Since the ordering wave vectors ${\bm Q}_\pm$ are incommensurate, the near-nodal Bogoliubov quasiparticles do not satisfy this requirement and give a small contribution to this coefficient. In contrast, once the pockets grow to include areas away from the nodes, near $(\pm \pi, 0)$ and $(0,\pm \pi)$ points, the constraint is satisfied. Therefore the coefficient of the $|m_{{\bm Q}_+}|^2|m_{{\bm Q}_-}|^2$ is small at low fields, but, as the field increases and the pockets grow, the two components of the magnetization
begin to interfere destructively, leading to an increase in the energy of the 2Q phase relative to that of the 1Q phase. To test this argument we computed the average occupation number of Bogoliubov quasiparticles with a given wave-vector ${\bm k}$ over
the entire Brillouin zone for the parameters where 2Q/1Q transition exists.
The results, shown in Fig. \ref{fig:nB},
clearly indicate that the 1Q phase becomes stable as soon as the overlap between particle-hole clouds from different pockets becomes strong.
Therefore, the main physical parameter that decides between the two possible scenarios shown in Figs. \ref{fig:PDs}(a) and (b) is the barrier between orthogonal pockets which, in turn,
depends on the shape of the Fermi surface. Recall that the SDW phase appears for values of the magnetic interaction $J\geq 0.85-0.9 J_c$, where $J_c$ is the critical value when the normal state becomes unstable towards magnetic order ~\cite{Kato11}; this result remains valid for any Fermi surface shape. We verified that, for a Fermi surface where 2Q-1Q transition exists, the 1Q phase persists over wide range of values of the pairing interaction $V$, although, once $V$ is significantly reduced, the 1Q region becomes too narrow to be found numerically in our calculation. At the same time, for the Fermi surface where only 2Q order is found, such as that used in the main panel of Fig. \ref{fig:PDs}(a), no changes in the interaction can induce the 1Q phase.

\begin{figure}[t]
\includegraphics[angle=0,width=8.5cm,trim= 60 40 40 10,clip]{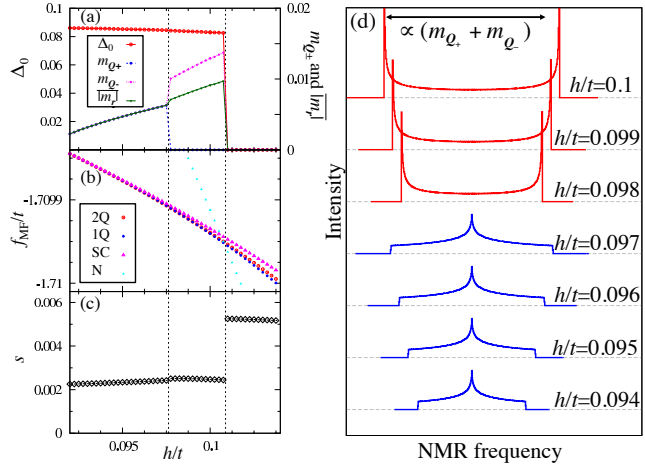}
\caption{Mean-field solutions for $t'/t=0.2$, $V/t=0.8$, $J/t=2.4$, and $T/t=0.003$ with $L=432$.
(a) Order parameters ($\Delta_0$, $m_{{\bm Q}_+}$, and $m_{{\bm Q}_-}$)
and the spatially-averaged magnetization $\overline{|m_{\bm r}|}$.
(b) Free energy densities $f_{\rm {MF}}/t$ for different solutions of self-consistent equations \eqref{eq:sceq2}.
(c) Entropy $s$ as a function of magnetic field.
(d) Expected NMR spectrum.
}
\label{fig:MFsols}
\end{figure}

Given that the 1Q and 2Q phases can only exist together in the phase diagram, we analyse the phase transition between the two.
Figures \ref{fig:MFsols}(a-c) show the solutions of Eqs.\eqref{eq:sceq1}-\eqref{eq:sceq2}  at low $T$ for the parameters in Fig. \ref{fig:PDs}(b).
While both the SC and the SDW order parameters show a significant jump at the transition from the normal paramagnetic phase, the change in $\Delta_0$ at the 1Q-2Q transition is extremely small, and the reduction in $m_{{\bm Q}_+}$ is compensated by the concomitant increase in $m_{{\bm Q}_-}$, so that the spatially-averaged magnetization
$\overline{|m_{\bm r}|}^2 \equiv (1/L^d)\sum_{\bm r} m^2_{\bm r} = (m^2_{{\bm Q}_+} + m^2_{{\bm Q}_-} )/2$,
remains nearly unchanged.
This is also clear from the almost unnoticeable kink in the free energy density at the transition line. As a result, the latent heat at the transition between N and 1Q phases  is much larger than the latent heat at the 1Q-2Q transition (see Fig.~\ref{fig:MFsols}(c)).
In other words, the 1Q-2Q transition, while first order, has very weak thermodynamic signatures, and is difficult to detect by  bulk measurements.
On the other hand, probes sensitive to the local magnetic structure, such as neutron scattering or NMR, should be able to clearly distinguish these two different phases.
Figure \ref{fig:MFsols}(d) shows the predicted NMR line shape in the magnetically ordered ${\bm Q}$-SDW phase and for different values of the magnetic field, obtained
by  calculating the  distribution of local magnetic field  generated by the  sum of the four Ce spins that surround each In(1) site~\cite{Koutroulakis10}.
There is a qualitative difference between the NMR line-shapes for the 1Q and 2Q phases:
while  the 1Q phase leads to a double-horn NMR line shape characteristic of 1D modulations, the line shape of the 2Q phase
has a maximum at the center (logarithmic Van-Hove singularity) typical of 2D modulations. The  NMR line shape as a function
of increasing field, see Fig.~\ref{fig:MFsols}(d), is very consistent with recent measurements from two different groups \cite{Koutroulakis10,Kumagai11}.






{\it Conclusions.} In summary, we presented a theory for the magnetic structure of the low-temperature and high field phase of Pauli limited nodal superconductors such as CeCoIn$_5$. Our main finding is that such materials are expected to broadly fall into two classes depending predominantly on the shape of the underlying Fermi surface: those exhibiting only the phase in which the local moments are modulated along the two orthogonal directions connecting opposite nodes of the superconducting gap, and those exhibiting a transition between such a doubly modulated phase and SDW order along a single inter-nodal direction. The origin of the transition is in the destructive interference of  scattering processes of Bogoliubov quasiparticles with two orthogonal wave-vectors. Thermodynamic signatures of the 2Q-1Q transition are very weak, and therefore local magnetic probes are best suited for probing it. The NMR line shape in our analysis is in a good agreement with recent data. It would be highly desirable to further test the proposed transition using neutron scattering.

{\it Acknowledgements}:
I. V. acknowledges support from NSF Grant No. DMR-1105339. Work at LANL was performed under the auspices of the
U.S.\ DOE contract No.~DE-AC52-06NA25396 through the LDRD program.


\begin{thebibliography}{25}
\expandafter\ifx\csname natexlab\endcsname\relax\def\natexlab#1{#1}\fi
\expandafter\ifx\csname bibnamefont\endcsname\relax
  \def\bibnamefont#1{#1}\fi
\expandafter\ifx\csname bibfnamefont\endcsname\relax
  \def\bibfnamefont#1{#1}\fi
\expandafter\ifx\csname citenamefont\endcsname\relax
  \def\citenamefont#1{#1}\fi
\expandafter\ifx\csname url\endcsname\relax
  \def\url#1{\texttt{#1}}\fi
\expandafter\ifx\csname urlprefix\endcsname\relax\def\urlprefix{URL }\fi
\providecommand{\bibinfo}[2]{#2}
\providecommand{\eprint}[2][]{\url{#2}}

\bibitem[{\citenamefont{Bianchi
  et~al.}(2003{\natexlab{a}})\citenamefont{Bianchi, Movshovich, Capan,
  Pagliuso, and Sarrao}}]{ABianchi:2003a}
\bibinfo{author}{\bibfnamefont{A.}~\bibnamefont{Bianchi}},
  \bibinfo{author}{\bibfnamefont{R.}~\bibnamefont{Movshovich}},
  \bibinfo{author}{\bibfnamefont{C.}~\bibnamefont{Capan}},
  \bibinfo{author}{\bibfnamefont{P.~G.} \bibnamefont{Pagliuso}},
  \bibnamefont{and} \bibinfo{author}{\bibfnamefont{J.~L.}
  \bibnamefont{Sarrao}}, \bibinfo{journal}{Phys. Rev. Lett.}
  \textbf{\bibinfo{volume}{91}}, \bibinfo{pages}{187004}
  (\bibinfo{year}{2003}{\natexlab{a}}).

\bibitem[{\citenamefont{Mitrovi{\'c} et~al.}(2006)\citenamefont{Mitrovi{\'c},
  Horvati{\'c}, Berthier, Knebel, Lapertot, and Flouquet}}]{VMitrovic:2006}
\bibinfo{author}{\bibfnamefont{V.~F.} \bibnamefont{Mitrovi{\'c}}},
  \bibinfo{author}{\bibfnamefont{M.}~\bibnamefont{Horvati{\'c}}},
  \bibinfo{author}{\bibfnamefont{C.}~\bibnamefont{Berthier}},
  \bibinfo{author}{\bibfnamefont{G.}~\bibnamefont{Knebel}},
  \bibinfo{author}{\bibfnamefont{G.}~\bibnamefont{Lapertot}}, \bibnamefont{and}
  \bibinfo{author}{\bibfnamefont{J.}~\bibnamefont{Flouquet}},
  \bibinfo{journal}{Phys. Rev. Lett.} \textbf{\bibinfo{volume}{97}},
  \bibinfo{pages}{117002} (\bibinfo{year}{2006}).

\bibitem[{\citenamefont{Young et~al.}(2007)\citenamefont{Young, Urbano, Curro,
  Thompson, Sarrao, Vorontsov, and Graf}}]{BLYoung:2007}
\bibinfo{author}{\bibfnamefont{B.-L.} \bibnamefont{Young}},
  \bibinfo{author}{\bibfnamefont{R.~R.} \bibnamefont{Urbano}},
  \bibinfo{author}{\bibfnamefont{N.~J.} \bibnamefont{Curro}},
  \bibinfo{author}{\bibfnamefont{J.~D.} \bibnamefont{Thompson}},
  \bibinfo{author}{\bibfnamefont{J.~L.} \bibnamefont{Sarrao}},
  \bibinfo{author}{\bibfnamefont{A.~B.} \bibnamefont{Vorontsov}},
  \bibnamefont{and} \bibinfo{author}{\bibfnamefont{M.~J.} \bibnamefont{Graf}},
  \bibinfo{journal}{Phys. Rev. Lett.} \textbf{\bibinfo{volume}{98}},
  \bibinfo{pages}{036402} (\bibinfo{year}{2007}).

\bibitem[{\citenamefont{Kenzelmann
  et~al.}(2008{\natexlab{a}})\citenamefont{Kenzelmann, Str{\"a}ssle,
  Niedermayer, Sigrist, Padmanabhan, Zolliker, Bianchi, Movshovich, Bauer,
  Sarrao et~al.}}]{MKenzelmann:2008}
\bibinfo{author}{\bibfnamefont{M.}~\bibnamefont{Kenzelmann}},
  \bibinfo{author}{\bibfnamefont{T.}~\bibnamefont{Str{\"a}ssle}},
  \bibinfo{author}{\bibfnamefont{C.}~\bibnamefont{Niedermayer}},
  \bibinfo{author}{\bibfnamefont{M.}~\bibnamefont{Sigrist}},
  \bibinfo{author}{\bibfnamefont{B.}~\bibnamefont{Padmanabhan}},
  \bibinfo{author}{\bibfnamefont{M.}~\bibnamefont{Zolliker}},
  \bibinfo{author}{\bibfnamefont{A.~D.} \bibnamefont{Bianchi}},
  \bibinfo{author}{\bibfnamefont{R.}~\bibnamefont{Movshovich}},
  \bibinfo{author}{\bibfnamefont{E.~D.} \bibnamefont{Bauer}},
  \bibinfo{author}{\bibfnamefont{J.~L.} \bibnamefont{Sarrao}},
  \bibnamefont{et~al.}, \bibinfo{journal}{Science}
  \textbf{\bibinfo{volume}{321}}, \bibinfo{pages}{1652}
  (\bibinfo{year}{2008}{\natexlab{a}}).

\bibitem[{\citenamefont{Koutroulakis
  et~al.}(2010{\natexlab{a}})\citenamefont{Koutroulakis, Stewart,
  Mitrovi{\'{c}}, Horvati{\'{c}}, Berthier, Lapertot, and
  Flouquet}}]{Koutroulakis10}
\bibinfo{author}{\bibfnamefont{G.}~\bibnamefont{Koutroulakis}},
  \bibinfo{author}{\bibfnamefont{M.~D.} \bibnamefont{Stewart}},
  \bibinfo{author}{\bibfnamefont{V.~F.} \bibnamefont{Mitrovi{\'{c}}}},
  \bibinfo{author}{\bibfnamefont{M.}~\bibnamefont{Horvati{\'{c}}}},
  \bibinfo{author}{\bibfnamefont{C.}~\bibnamefont{Berthier}},
  \bibinfo{author}{\bibfnamefont{G.}~\bibnamefont{Lapertot}}, \bibnamefont{and}
  \bibinfo{author}{\bibfnamefont{J.}~\bibnamefont{Flouquet}},
  \bibinfo{journal}{Phys. Rev. Lett.} \textbf{\bibinfo{volume}{104}},
  \bibinfo{pages}{087001} (\bibinfo{year}{2010}{\natexlab{a}}).

\bibitem[{\citenamefont{Kenzelmann et~al.}(2010)\citenamefont{Kenzelmann,
  Gerber, Egetenmeyer, Gavilano, Str\"assle, Bianchi, Ressouche, Movshovich,
  Bauer, Sarrao et~al.}}]{MKenzelmann:2010}
\bibinfo{author}{\bibfnamefont{M.}~\bibnamefont{Kenzelmann}},
  \bibinfo{author}{\bibfnamefont{S.}~\bibnamefont{Gerber}},
  \bibinfo{author}{\bibfnamefont{N.}~\bibnamefont{Egetenmeyer}},
  \bibinfo{author}{\bibfnamefont{J.~L.} \bibnamefont{Gavilano}},
  \bibinfo{author}{\bibfnamefont{T.}~\bibnamefont{Str\"assle}},
  \bibinfo{author}{\bibfnamefont{A.~D.} \bibnamefont{Bianchi}},
  \bibinfo{author}{\bibfnamefont{E.}~\bibnamefont{Ressouche}},
  \bibinfo{author}{\bibfnamefont{R.}~\bibnamefont{Movshovich}},
  \bibinfo{author}{\bibfnamefont{E.~D.} \bibnamefont{Bauer}},
  \bibinfo{author}{\bibfnamefont{J.~L.} \bibnamefont{Sarrao}},
  \bibnamefont{et~al.}, \bibinfo{journal}{Phys. Rev. Lett.}
  \textbf{\bibinfo{volume}{104}}, \bibinfo{pages}{127001}
  (\bibinfo{year}{2010}).

\bibitem[{\citenamefont{Blackburn et~al.}(2010)\citenamefont{Blackburn, Das,
  Eskildsen, Forgan, Laver, Niedermayer, Petrovic, and
  White}}]{EBlackburn:2010}
\bibinfo{author}{\bibfnamefont{E.}~\bibnamefont{Blackburn}},
  \bibinfo{author}{\bibfnamefont{P.}~\bibnamefont{Das}},
  \bibinfo{author}{\bibfnamefont{M.~R.} \bibnamefont{Eskildsen}},
  \bibinfo{author}{\bibfnamefont{E.~M.} \bibnamefont{Forgan}},
  \bibinfo{author}{\bibfnamefont{M.}~\bibnamefont{Laver}},
  \bibinfo{author}{\bibfnamefont{C.}~\bibnamefont{Niedermayer}},
  \bibinfo{author}{\bibfnamefont{C.}~\bibnamefont{Petrovic}}, \bibnamefont{and}
  \bibinfo{author}{\bibfnamefont{J.~S.} \bibnamefont{White}},
  \bibinfo{journal}{Phys. Rev. Lett.} \textbf{\bibinfo{volume}{105}},
  \bibinfo{pages}{187001} (\bibinfo{year}{2010}).

\bibitem[{\citenamefont{Bianchi
  et~al.}(2003{\natexlab{b}})\citenamefont{Bianchi, Movshovich, Vekhter,
  Pagliuso, and Sarrao}}]{ABianchi:2003b}
\bibinfo{author}{\bibfnamefont{A.}~\bibnamefont{Bianchi}},
  \bibinfo{author}{\bibfnamefont{R.}~\bibnamefont{Movshovich}},
  \bibinfo{author}{\bibfnamefont{I.}~\bibnamefont{Vekhter}},
  \bibinfo{author}{\bibfnamefont{P.~G.} \bibnamefont{Pagliuso}},
  \bibnamefont{and} \bibinfo{author}{\bibfnamefont{J.~L.}
  \bibnamefont{Sarrao}}, \bibinfo{journal}{Phys. Rev. Lett.}
  \textbf{\bibinfo{volume}{91}}, \bibinfo{pages}{257001}
  (\bibinfo{year}{2003}{\natexlab{b}}).

\bibitem[{\citenamefont{Bauer et~al.}(2005)\citenamefont{Bauer, Capan, Ronning,
  Movshovich, Thompson, and Sarrao}}]{EDBauer:2005}
\bibinfo{author}{\bibfnamefont{E.~D.} \bibnamefont{Bauer}},
  \bibinfo{author}{\bibfnamefont{C.}~\bibnamefont{Capan}},
  \bibinfo{author}{\bibfnamefont{F.}~\bibnamefont{Ronning}},
  \bibinfo{author}{\bibfnamefont{R.}~\bibnamefont{Movshovich}},
  \bibinfo{author}{\bibfnamefont{J.~D.} \bibnamefont{Thompson}},
  \bibnamefont{and} \bibinfo{author}{\bibfnamefont{J.~L.}
  \bibnamefont{Sarrao}}, \bibinfo{journal}{Phys. Rev. Lett.}
  \textbf{\bibinfo{volume}{94}}, \bibinfo{pages}{047001}
  (\bibinfo{year}{2005}).

\bibitem[{\citenamefont{Ronning et~al.}(2006)\citenamefont{Ronning, Capan,
  Bauer, Thompson, Sarrao, and Movshovich}}]{FRonning:2006}
\bibinfo{author}{\bibfnamefont{F.}~\bibnamefont{Ronning}},
  \bibinfo{author}{\bibfnamefont{C.}~\bibnamefont{Capan}},
  \bibinfo{author}{\bibfnamefont{E.~D.} \bibnamefont{Bauer}},
  \bibinfo{author}{\bibfnamefont{J.~D.} \bibnamefont{Thompson}},
  \bibinfo{author}{\bibfnamefont{J.~L.} \bibnamefont{Sarrao}},
  \bibnamefont{and}
  \bibinfo{author}{\bibfnamefont{R.}~\bibnamefont{Movshovich}},
  \bibinfo{journal}{Phys. Rev. B} \textbf{\bibinfo{volume}{73}},
  \bibinfo{pages}{064519} (\bibinfo{year}{2006}).

\bibitem[{\citenamefont{Miclea et~al.}(2006)\citenamefont{Miclea, Nicklas,
  Parker, Maki, Sarrao, Thompson, Sparn, and Steglich}}]{CFMiclea:2006}
\bibinfo{author}{\bibfnamefont{C.~F.} \bibnamefont{Miclea}},
  \bibinfo{author}{\bibfnamefont{M.}~\bibnamefont{Nicklas}},
  \bibinfo{author}{\bibfnamefont{D.}~\bibnamefont{Parker}},
  \bibinfo{author}{\bibfnamefont{K.}~\bibnamefont{Maki}},
  \bibinfo{author}{\bibfnamefont{J.~L.} \bibnamefont{Sarrao}},
  \bibinfo{author}{\bibfnamefont{J.~D.} \bibnamefont{Thompson}},
  \bibinfo{author}{\bibfnamefont{G.}~\bibnamefont{Sparn}}, \bibnamefont{and}
  \bibinfo{author}{\bibfnamefont{F.}~\bibnamefont{Steglich}},
  \bibinfo{journal}{Phys. Rev. Lett.} \textbf{\bibinfo{volume}{96}},
  \bibinfo{pages}{117001} (\bibinfo{year}{2006}).

\bibitem[{\citenamefont{Agterberg et~al.}(2009)\citenamefont{Agterberg,
  Sigrist, and Tsunetsugu}}]{DFAgterberg:2009}
\bibinfo{author}{\bibfnamefont{D.~F.} \bibnamefont{Agterberg}},
  \bibinfo{author}{\bibfnamefont{M.}~\bibnamefont{Sigrist}}, \bibnamefont{and}
  \bibinfo{author}{\bibfnamefont{H.}~\bibnamefont{Tsunetsugu}},
  \bibinfo{journal}{Phys. Rev. Lett.} \textbf{\bibinfo{volume}{102}},
  \bibinfo{pages}{207004} (\bibinfo{year}{2009}).

\bibitem[{\citenamefont{Yanase and Sigrist}(2009)}]{YYanase:2009}
\bibinfo{author}{\bibfnamefont{Y.}~\bibnamefont{Yanase}} \bibnamefont{and}
  \bibinfo{author}{\bibfnamefont{M.}~\bibnamefont{Sigrist}},
  \bibinfo{journal}{Journal of the Physical Society of Japan}
  \textbf{\bibinfo{volume}{78}}, \bibinfo{pages}{114715}
  (\bibinfo{year}{2009}),

\bibitem[{\citenamefont{Yanase and Sigrist}(2011{\natexlab{a}})}]{Yanase2011}
\bibinfo{author}{\bibfnamefont{Y.}~\bibnamefont{Yanase}} \bibnamefont{and}
  \bibinfo{author}{\bibfnamefont{M.}~\bibnamefont{Sigrist}},
  \bibinfo{journal}{Journal of the Physical Society of Japan}
  \textbf{\bibinfo{volume}{80}}, \bibinfo{pages}{094702}
  (\bibinfo{year}{2011}{\natexlab{a}}), 

\bibitem[{\citenamefont{Yanase and Sigrist}(2011{\natexlab{b}})}]{Yanase2011a}
\bibinfo{author}{\bibfnamefont{Y.}~\bibnamefont{Yanase}} \bibnamefont{and}
  \bibinfo{author}{\bibfnamefont{M.}~\bibnamefont{Sigrist}},
  \bibinfo{journal}{J. Phys.: Condensed matter} \textbf{\bibinfo{volume}{23}}, \bibinfo{pages}{094219}
  (\bibinfo{year}{2011}{\natexlab{b}}), 

\bibitem[{\citenamefont{Kumagai et~al.}(2011)\citenamefont{Kumagai, Shishido,
  Shibauchi, and Matsuda}}]{Kumagai11}
\bibinfo{author}{\bibfnamefont{K.}~\bibnamefont{Kumagai}},
  \bibinfo{author}{\bibfnamefont{H.}~\bibnamefont{Shishido}},
  \bibinfo{author}{\bibfnamefont{T.}~\bibnamefont{Shibauchi}},
  \bibnamefont{and} \bibinfo{author}{\bibfnamefont{Y.}~\bibnamefont{Matsuda}},
  \bibinfo{journal}{Phys. Rev. Lett.} \textbf{\bibinfo{volume}{106}},
  \bibinfo{pages}{137004} (\bibinfo{year}{2011}),

\bibitem[{\citenamefont{Aperis et~al.}(2008)\citenamefont{Aperis,
  Varelogiannis, Littlewood, and Simons}}]{AAperis:2008}
\bibinfo{author}{\bibfnamefont{A.}~\bibnamefont{Aperis}},
  \bibinfo{author}{\bibfnamefont{G.}~\bibnamefont{Varelogiannis}},
  \bibinfo{author}{\bibfnamefont{P.~B.} \bibnamefont{Littlewood}},
  \bibnamefont{and} \bibinfo{author}{\bibfnamefont{B.~D.}
  \bibnamefont{Simons}}, \bibinfo{journal}{Journal of Physics: Condensed
  Matter} \textbf{\bibinfo{volume}{20}}, \bibinfo{pages}{434235}
  (\bibinfo{year}{2008}).

\bibitem[{\citenamefont{Aperis et~al.}(2010)\citenamefont{Aperis,
  Varelogiannis, and Littlewood}}]{AAperis:2010}
\bibinfo{author}{\bibfnamefont{A.}~\bibnamefont{Aperis}},
  \bibinfo{author}{\bibfnamefont{G.}~\bibnamefont{Varelogiannis}},
  \bibnamefont{and} \bibinfo{author}{\bibfnamefont{P.~B.}
  \bibnamefont{Littlewood}}, \bibinfo{journal}{Phys. Rev. Lett.}
  \textbf{\bibinfo{volume}{104}}, \bibinfo{pages}{216403}
  (\bibinfo{year}{2010}).

\bibitem[{\citenamefont{Ikeda et~al.}(2010)\citenamefont{Ikeda, Hatakeyama, and
  Aoyama}}]{RIkeda:2010}
\bibinfo{author}{\bibfnamefont{R.}~\bibnamefont{Ikeda}},
  \bibinfo{author}{\bibfnamefont{Y.}~\bibnamefont{Hatakeyama}},
  \bibnamefont{and} \bibinfo{author}{\bibfnamefont{K.}~\bibnamefont{Aoyama}},
  \bibinfo{journal}{Phys. Rev. B} \textbf{\bibinfo{volume}{82}},
  \bibinfo{pages}{060510} (\bibinfo{year}{2010}).

\bibitem[{\citenamefont{Suzuki et~al.}(2011)\citenamefont{Suzuki, Ichioka, and
  Machida}}]{KSuzuki:2011}
\bibinfo{author}{\bibfnamefont{K.}~\bibnamefont{Suzuki}},
  \bibinfo{author}{\bibfnamefont{M.}~\bibnamefont{Ichioka}}, \bibnamefont{and}
  \bibinfo{author}{\bibfnamefont{K.}~\bibnamefont{Machida}},
  \bibinfo{journal}{Physical Review B} \textbf{\bibinfo{volume}{83}},
  \bibinfo{pages}{140503} (\bibinfo{year}{2011}).

\bibitem[{\citenamefont{Kato et~al.}(2011)\citenamefont{Kato, Batista, and
  Vekhter}}]{Kato11}
\bibinfo{author}{\bibfnamefont{Y.}~\bibnamefont{Kato}},
  \bibinfo{author}{\bibfnamefont{C.~D.} \bibnamefont{Batista}},
  \bibnamefont{and} \bibinfo{author}{\bibfnamefont{I.}~\bibnamefont{Vekhter}},
  \bibinfo{journal}{Phys. Rev. Lett.} \textbf{\bibinfo{volume}{107}},
  \bibinfo{pages}{096401} (\bibinfo{year}{2011}).

\bibitem[{\citenamefont{Curro et~al.}(2009)\citenamefont{Curro, Young, Urbano,
  and Graf}}]{Curro2009}
\bibinfo{author}{\bibfnamefont{N.~J.} \bibnamefont{Curro}},
  \bibinfo{author}{\bibfnamefont{B.-L.} \bibnamefont{Young}},
  \bibinfo{author}{\bibfnamefont{R.~R.} \bibnamefont{Urbano}},
  \bibnamefont{and} \bibinfo{author}{\bibfnamefont{M.~J.} \bibnamefont{Graf}},
  \bibinfo{journal}{Journal of Low Temperature Physics}
  \textbf{\bibinfo{volume}{158}}, \bibinfo{pages}{635} (\bibinfo{year}{2009}),

\bibitem[{\citenamefont{Kenzelmann
  et~al.}(2008{\natexlab{b}})\citenamefont{Kenzelmann, Str\"assle, Niedermayer,
  Sigrist, Padmanabhan, Zolliker, Bianchi, Movshovich, Bauer, Sarrao
  et~al.}}]{Kenzelmann08}
\bibinfo{author}{\bibfnamefont{M.}~\bibnamefont{Kenzelmann}},
  \bibinfo{author}{\bibfnamefont{T.}~\bibnamefont{Str\"assle}},
  \bibinfo{author}{\bibfnamefont{C.}~\bibnamefont{Niedermayer}},
  \bibinfo{author}{\bibfnamefont{M.}~\bibnamefont{Sigrist}},
  \bibinfo{author}{\bibfnamefont{B.}~\bibnamefont{Padmanabhan}},
  \bibinfo{author}{\bibfnamefont{M.}~\bibnamefont{Zolliker}},
  \bibinfo{author}{\bibfnamefont{A.~D.} \bibnamefont{Bianchi}},
  \bibinfo{author}{\bibfnamefont{R.}~\bibnamefont{Movshovich}},
  \bibinfo{author}{\bibfnamefont{E.~D.} \bibnamefont{Bauer}},
  \bibinfo{author}{\bibfnamefont{J.~L.} \bibnamefont{Sarrao}},
  \bibnamefont{et~al.}, \bibinfo{journal}{Science}
  \textbf{\bibinfo{volume}{321}}, \bibinfo{pages}{1652}
  (\bibinfo{year}{2008}{\natexlab{b}}),

\bibitem[{\citenamefont{Yang and Sondhi}(1998)}]{Yang98}
\bibinfo{author}{\bibfnamefont{K.}~\bibnamefont{Yang}} \bibnamefont{and}
  \bibinfo{author}{\bibfnamefont{S.~L.} \bibnamefont{Sondhi}},
  \bibinfo{journal}{Phys. Rev. B} \textbf{\bibinfo{volume}{57}},
  \bibinfo{pages}{8566} (\bibinfo{year}{1998}),



\end{thebibliography}

\end{document}